\documentclass[twocolumn,prl,twocolumn,superscriptaddress]{revtex4}
\usepackage{color}
\usepackage{amsmath}
\usepackage{amssymb}
\usepackage{graphicx}
\usepackage{tikz}
\usepackage{physics}

\usepackage{float}
\usepackage{bbm}

\usepackage{epsfig} 
\usepackage{verbatim}
\usepackage{array}
\usepackage{setspace}

\usepackage[unicode=true,
 bookmarks=true,bookmarksnumbered=false,bookmarksopen=false,
 breaklinks=false,pdfborder={0 0 1},backref=false,colorlinks=true]
 {hyperref}
\hypersetup{
 linkcolor=magenta, urlcolor=blue, citecolor=blue, pdfstartview={FitH}, hyperfootnotes=false, unicode=true}
 
\usepackage{color}
\newcommand*{\white}{\textcolor{white}} 

\makeatletter
\@ifundefined{textcolor}{}
{%
 \definecolor{BLACK}{gray}{0}
 \definecolor{WHITE}{gray}{1}
 \definecolor{RED}{rgb}{1,0,0}
 \definecolor{GREEN}{rgb}{0,1,0}
 \definecolor{BLUE}{rgb}{0,0,1}
 \definecolor{CYAN}{cmyk}{1,0,0,0}
 \definecolor{MAGENTA}{cmyk}{0,1,0,0}
 \definecolor{YELLOW}{cmyk}{0,0,1,0}
}

\usepackage{amsfonts}\usepackage{tabularx}\usepackage{dcolumn}\usepackage{bm}\usepackage{graphicx}\usepackage{epstopdf}

\hypersetup{urlcolor=blue}

\makeatother

\newcolumntype{C}[1]{>{\centering\arraybackslash$}p{#1}<{$}}

\usepackage{algorithm}
\usepackage{algorithmic}

\begin{document}

\widetext

\title{When does reinforcement learning stand out in quantum control? A comparative study on state preparation}

\author{Xiao-Ming Zhang}
\affiliation{Department of Physics, City University of Hong Kong, Tat Chee Avenue, Kowloon, Hong Kong SAR, China}
\affiliation{Shenzhen Research Institute, City University of Hong Kong, Shenzhen, Guangdong 518057, China}

\author{Zezhu Wei}
\affiliation{Department of Physics, City University of Hong Kong, Tat Chee Avenue, Kowloon, Hong Kong SAR, China}

\author{Raza Asad}
\affiliation{Department of Mathematics, City University of Hong Kong, Tat Chee Avenue, Kowloon, Hong Kong SAR, China}

\author{Xu-Chen Yang}
\affiliation{Department of Physics, The University of Hong Kong, Pokfulam, Hong Kong SAR, China}

\author{Xin Wang}
\affiliation{Department of Physics, City University of Hong Kong, Tat Chee Avenue, Kowloon, Hong Kong SAR, China}
\affiliation{Shenzhen Research Institute, City University of Hong Kong, Shenzhen, Guangdong 518057, China}
\email{x.wang@cityu.edu.hk}

\begin{abstract}
Reinforcement learning has been widely used in many problems, including quantum control of qubits. However, such problems can, at the same time, be solved by traditional, non-machine-learning methods, such as stochastic gradient descent and Krotov algorithms, and it remains unclear which one is most suitable when the control has specific constraints. In this work, we perform a comparative study on the efficacy of three reinforcement learning algorithms: tabular Q-learning, deep Q-learning, and policy gradient, as well as two non-machine-learning methods: stochastic gradient descent and Krotov algorithms, in the problem of preparing a desired quantum state. 
We found that overall, the deep Q-learning and policy gradient algorithms outperform others when the problem is discretized, e.g. allowing discrete values of control, and when the problem scales up. The reinforcement learning algorithms
can also adaptively reduce the complexity of the control sequences, shortening the operation time and improving the fidelity. Our comparison provides insights into the suitability of reinforcement learning in quantum control problems.
\end{abstract}

\maketitle

\section{Introduction}
Reinforcement learning, a branch of machine learning in artificial intelligence, has proven to be a powerful tool to solve a wide range of complex problems,
such as the games of Go \cite{Silver.17} and Atari \cite{Mnih.15}. Reinforcement learning has also been applied to a variety of problems in quantum physics with vast success \cite{Chen.13,Chen.19,Bukov.18,Bukov.18(2),August.18,Albarran.18,Zhang.18,Niu.18,Fosel.18,Melnikov.18,Dunjko.16,Hentschel.14,Day.19,Wu.18,Ferrie.14,Reddy.16,Colabrese.18,Nautrup.18,Sweke.18,Andreasson.18,Halverson.19,Zhao.19}, including quantum state preparation \cite{Chen.13,Chen.19,Bukov.18,Bukov.18(2),August.18,Albarran.18}, state transfer \cite{Zhang.18}, quantum gate design \cite{Niu.18}, and error correction \cite{Fosel.18}. In many cases, it outperforms commonly-used conventional algorithms, such as Krotov and Stochastic Gradient Descent (SGD) algorithms \cite{Zhang.18,Niu.18}.
In the reinforcement learning algorithm, an optimization problem is converted to a set of policies that governs the behavior of a computer agent, i.e. its choices of actions and, consequently, the reward it receives. By simulating sequences of actions taken by the agent maximizing the reward, one finds an optimal solution to the desired problem \cite{Sutton}.

The development of techniques that efficiently optimize control protocols is key to quantum physics. While some problems can be solved analytically using methods such as reverse engineering \cite{Barnes.12}, in most cases numerical solutions are required. Various numerical methods are therefore put forward, such as gradient-based methods (including SGD \cite{Ferrie.14}, GRAPE \cite{Khaneja.05} and variants  \cite{Jager.14}), the Krotov method \cite{Krotov.96},  the  Nelder-Mead method \cite{Kelly.14} and convex programming \cite{Kosut.13}. Recently, there is a frenetic attempt to apply reinforcement learning and other machine-learning-based algorithms \cite{Wang.16,Carleo.17, Deng.17,Carrasquilla.17,Lijun.17,Hsu.18} to a wide range of physics problems. In particular, the introduction of reinforcement learning to quantum control  have revealed new interesting physics \cite{Chen.13,Chen.19,Bukov.18,Bukov.18(2),August.18,Albarran.18}, and these techniques have therefore received increasing attention. A fundamental question then arises: under what situation is reinforcement learning the most suitable method? 

In this paper, we consider problems related to quantum control of a qubit. The goal of these problems is typically to steer the qubit toward a target state under certain constraints. The mismatch between the final qubit state and the target state naturally serves as the cost function used in the SGD or Krotov methods, and the negative cost function can serve as the reward function in the reinforcement learning procedure. Our question then becomes: under different scenarios of constraints, which algorithm works best? In this work, we compare the efficacy of two commonly-used traditional methods: SGD and the Krotov method, and three algorithms based on reinforcement learning: tabular Q-learning (TQL) \cite{Sutton}, deep Q-learning (DQL) \cite{Mnih.15}, and policy gradient (PG) \cite{Sutton.00} , under situations with different types of control constraints.

In Ref.~\cite{Bukov.18}, the Q-learning techniques (TQL and DQL) have been applied to the problem of quantum state preparation, revealing different stages of quantum control.
The problem of preparing a desired quantum state from a given initial state is on one hand simple enough to be investigated in full detail, and on the other hand contains sufficient physics allowing for various types of control constraints. We therefore take quantum state preparation as the platform that our comparison of different algorithms is based on. While a detailed description of quantum state preparation is provided in Results, we briefly introduce the five algorithms we are comparing in this work here. (Detailed implementations are provided in the Methods and Supplementary Method 1.)

 \begin{figure}
\centering
\includegraphics[width=0.8 \columnwidth]{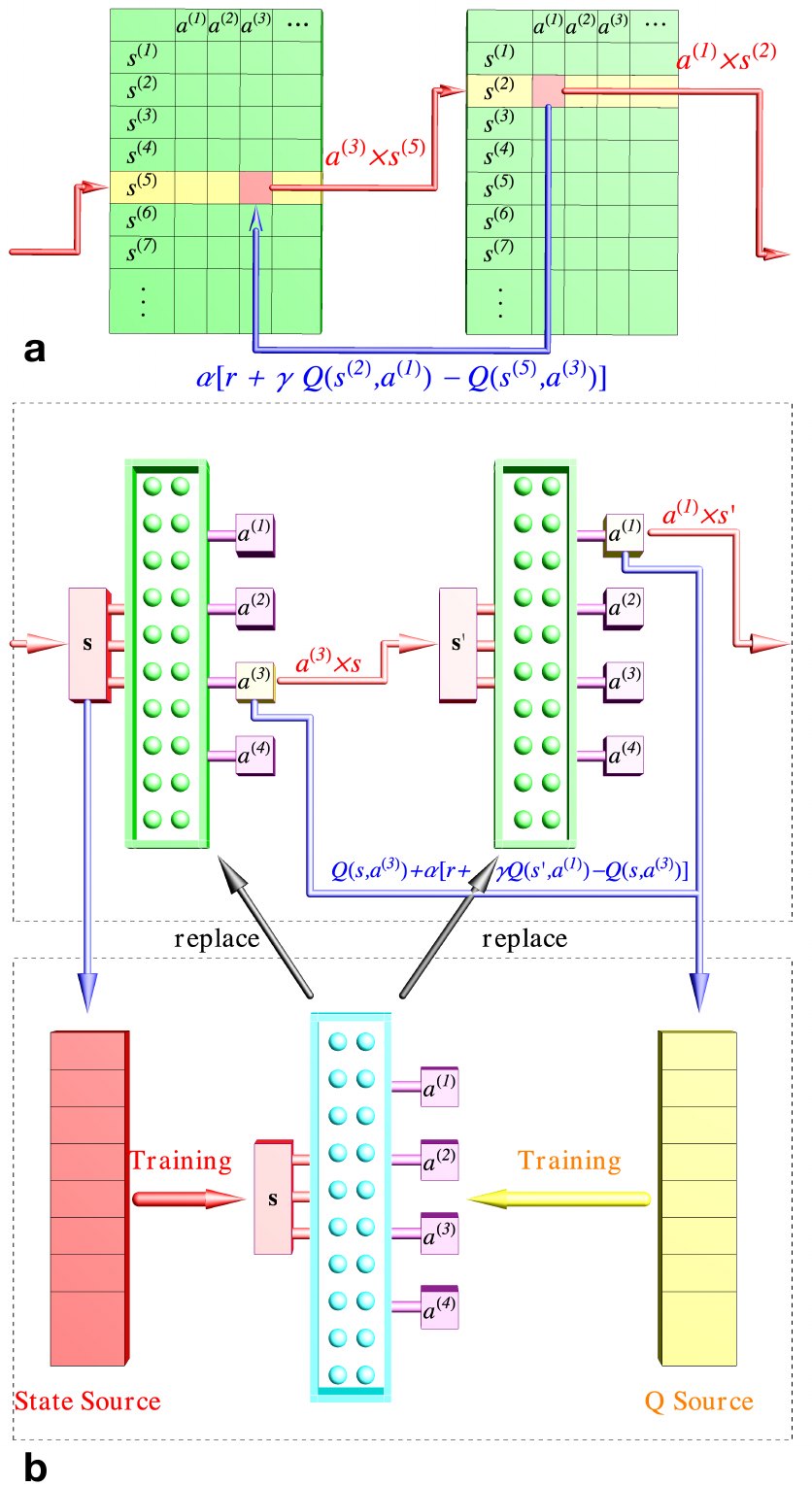}
\caption{ \textbf{Sketch of the procedure of TQL and DQL.}  (a) In TQL, the $Q(s,a)$ values are stored in the Q-table. When the agent is at state $s^{(5)}$, it reviews $Q(s^{(5)},a^{(i)})$ for all possible actions and chooses one with the maximum ``Q-value'' (which we assume is $a^{(3)}$). As a result, the state then evolves to $s^{(2)}$. Depending on the distance between $s^{(2)}$ and the target, the Q-values (e.g. $Q(s^{(5)},a^{(3)})$) is updated according to Eq.~\eqref{eq:q1}. This process is then repeated at the new state $s^{(2)}$ and so forth. (b) In DQL, the Q-table is replaced by the Q-network. Instead of choosing an action with the maximum Q-value from a list, this process is done by a neural network, the Q-network, which takes the input state $(\textbf{s})$ and outputs an action that it finds most appropriate. Evaluation of the resulting state $(\textbf{s}')$ after the action suggests how the neural network should be updated (trained). For detailed implementation, see Methods and Supplementary Method 1.\label{fig:schem}}
\end{figure}  
 
SGD is one of the simplest gradient-based optimization algorithms. In each iteration,  a direction in the parameter space is randomly chosen, along which the control field is updated using the gradient of the cost function defined as the mismatch between the evolved state and the target state. Ideally, the gradient is zero when the calculation has converged to the optimal solution. The Krotov algorithm has a different strategy: The initial state is first propagated forward obtaining the evolved state. The evolved state is then projected to the target state, defining a co-state encapsulating the mismatch between the two. Then the co-state is propagated backward to the initial state, during which process the control fields are updated. When the calculation is converged, the co-state is identical to the target state. 

In Q-learning (including TQL and DQL),  
a computer agent evolves in an environment.
All information required for optimization is encoded in the environment, which is allowed to be in a set of states $\mathcal{S}$. In each step, the agent chooses an action from a set $\mathcal{A}$, bringing the environment of the agent to another state. As a consequence, the agent acquires a reward, which encapsulates the desired optimization problem. Fig.~\ref{fig:schem}a schematically shows how TQL works. 
At each state $s\in\mathcal{S}$, the agent chooses actions $a\in\mathcal{A}$ according to the action-value function $Q(s,a)$, defined as the estimated total reward starting from state $s$ and action $a$, forming the so-called Q-table. Each time the agent takes an action, a reward $r$ is generated according to the distance between the resulting state and the target, which updates the Q-table. An optimal solution is found by iterating this process sufficient times.
We note that since a table has a finite number of entries, both the states and actions should be discretized. Fig.~\ref{fig:schem}b shows DQL, in which the role of the Q-table is replaced by a neural network, called the Q-network. The agent then chooses its action according to the output of the Q-network, and the reward is used to update the network. In this case, although the allowed actions are typically discrete, the input state can actually be continuous.

Similar to TQL and DQL, PG also requires the sets of states $\mathcal{S}$, actions  $\mathcal{A}$, and rewards $r$. The policy of the agent is represented by a neural network.  With the state as the input, the network outputs the probability of choosing each action. After each episode, the policy network is updated toward a direction that increases the total reward. Since the state is encoded as the input of the neural network, PG can also accommodate continuous input states.

\section{Results}
\subsection{Single-qubit case}
We start with the preparation of a single-qubit state. Consider the time dependent Hamiltonian
\begin{equation}
H[J(t)]=4 J(t) \sigma_{z} + h \sigma_{x},
\end{equation}
where $\sigma_{x}$ and $\sigma_{z}$ are Pauli matrices. The Hamiltonian may describe a singlet-triplet qubit \cite{Petta.05} or a single spin with energy gap $h$ under tunable control fields \cite{Greilich.09,Poem.11}. In these systems, it is difficult to vary $h$ during gate operations, and we therefore assume that $h$ is a constant in our work, which at the same time serves as our energy unit. Quantum control of the qubit is then achieved by altering $J(t)$ dynamically.

Quantum state preparation refers to the problem to find $J(t)$ such that a given initial state $|\psi_0\rangle$ evolves, within time $T$, to a final state $|\psi_\text{f}\rangle$ that is as close as possible to the target state $|\phi\rangle$. The quality of the state transfer is evaluated using the fidelity, defined as 
\begin{equation}
F=\left|\langle\psi_\text{f}|\phi\rangle\right|^2.
\label{eq:fid}
\end{equation}
We typically use the averaged fidelity $\overline{F}$ over many runs of a given algorithm in our comparison (unless otherwise noted, we average 100 runs to obtain $\overline{F}$), because the initial guesses of the control sequences are random, and the reinforcement learning procedure is probabilistic.

In this work, we take $|\psi_0\rangle=|0\rangle$, $|\phi\rangle=|1\rangle$ and $T=2\pi$ unless otherwise specified.  Under different situations, there are various kinds of preferences or restrictions of control. We consider the following types of restrictions:

 (i) Assuming that control is performed with a sequence of piecewise constant pulses, and in this work, we further assume that the time duration of each piece is equal to each other for convenience. For this purpose, we divide the total time $T$ into $N$ equal time steps,  each of which having a step size $dt=T/N$, with $N$ denoting the maximum number of pieces required by the control. $J(t)$ is accordingly discretized, so that on the $i$th time step, $J(t)=J_i$ and the system evolves under $H(J_i)$. Denoting the state at the end of the $i$th time step as $|\psi_i\rangle$, the evolution at the $i$th step is $|\psi_{i}\rangle=U_i|\psi_{i-1}\rangle$, where $U_i=\exp\{-i H(J_i) dt\}$. 
In principle, the evolution time can be less than $T$, namely the evolution may conclude at the $i_\text{f}$ th time step with $i_\text{f}\leqslant N$. (In our calculations, the evolution is terminated when the fidelity $F\ge0.999$.) Due to their nature, SGD and Krotov have to finish all time steps, i.e. $i_\text{f}=N$. However, as we shall see below, QL and DQL frequently have $i_\text{f}<N$.

(ii) We also consider the case where the magnitude of the control field is bounded, i.e. $J_{i}\in[J_{\max}, J_{\min}]$ for all $i$. The constraint can be straightforwardly satisfied in TQL and DQL, since they only operate within the given set of actions thus cannot exceed the bounds. For SGD and Krotov, updates to the control fields may exceed the bounds, in which case we need to enforce the bounds by setting $J_i$ as $J_{\max}$ when the updated value is greater than $J_{\max}$, and as $J_{\min}$ when the updated value is smaller than $J_{\min}$. In the case in which either of them is not restricted, we simply note $J_{\min}\rightarrow-\infty$ or $J_{\max}\rightarrow\infty$.
 
 (iii) The values of the control field may be discretized in the given range, i.e., $J_i\in\{J_{\min}, J_{\min} +\text{d}J/M, J_{\min}+2\text{d}J/M,\cdots, J_{\max}\}$ where $\text{d}J=(J_{\max}-J_{\min})$, so that the control field can take $M+1$ values including $J_{\min}$ and $J_{\max}$. In reality this situation may arise, for example, when decomposing a quantum operation into a set of given gates \cite{Kitaev.02,Harrow.02,Campbell.17}.
  For a reason similar to (ii), TQL and DQL only select actions within the given set so the constraint is satisfied. For SGD and Krotov which keep updating the values of the control field during iterations, we enforce the constraint by setting the value of each control field to the nearest allowed value at the end of the execution.

 \begin{figure}
\centering
\includegraphics[width= \columnwidth]{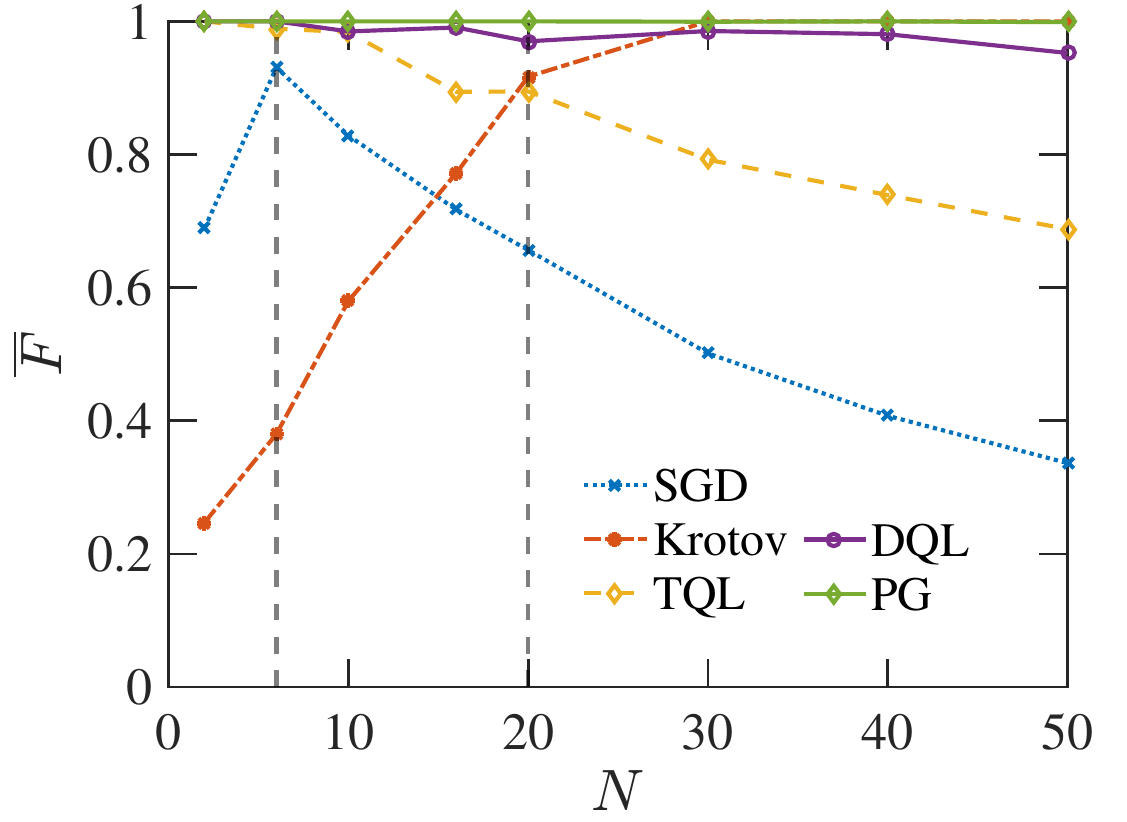}
\caption{ \textbf{Average fidelities as functions of the maximum number of control pieces.} For TQL, DQL and PG, $J_i\in\{0,1\}$ (i.e. $M=1$). For SGD and Krotov, no restriction is imposed on the range of $J_i$ and $M$ (i.e. $M\rightarrow\infty$). The vertical dashed lines correspond to results shown with respective $M$ values in Fig.~\ref{fig:Jrestrictedgen}. \label{fig:varyN}
}
\end{figure}

To sum up, the number of pieces in control sequences $N$, the bounds of the control field $J_{\min}$ and $J_{\max}$, as well as the number of the discrete values of the control field $M+1$ are the main factors characterizing situations to prepare quantum states, based on which our comparison of different algorithms is conducted. We also define $N^{\rm iter}$ as the number of iterations performed in executing an algorithm, which is typically taken as equal for different algorithms to ensure a fair comparison. Unless otherwise noted, $N^{\rm iter}=500$ in all results shown.

\begin{figure}
\centering
\includegraphics[width=0.95 \columnwidth]{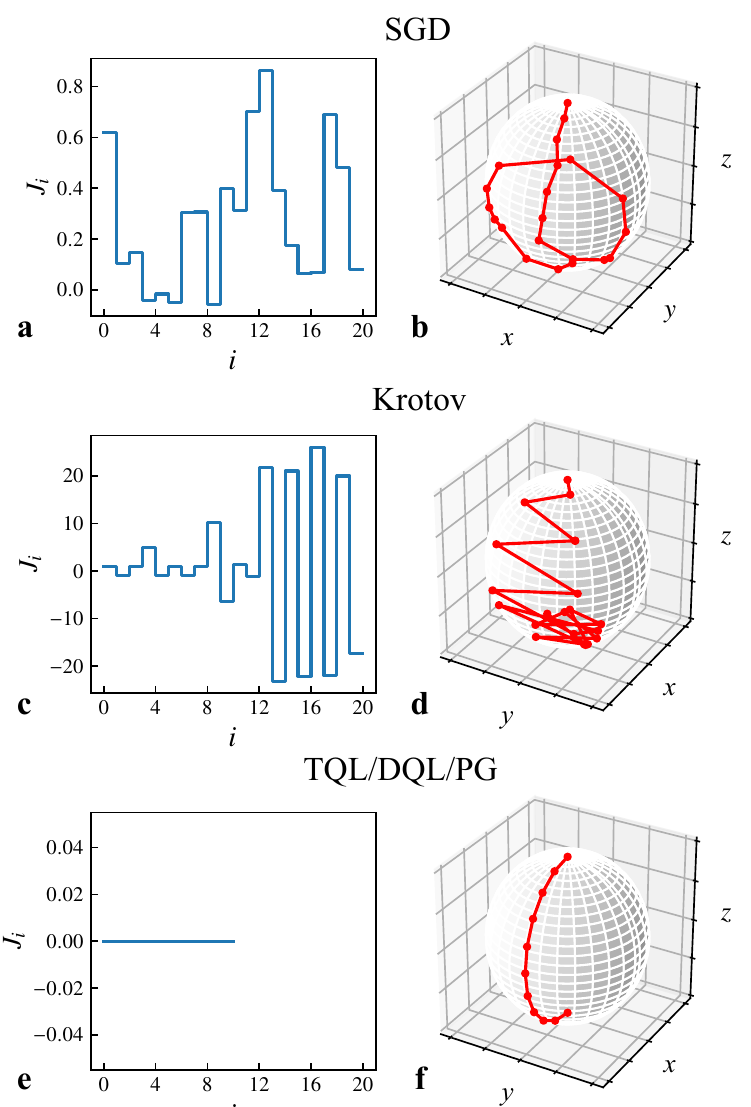}
\caption{\textbf{Pulse profiles and the corresponding trajectories on the Bloch sphere}. Left column: Example pulse profiles taken from results of Fig.~\ref{fig:varyN} with $N=20$. Right column: Evolution of the state corresponding to the respective control sequence in the left column. TQL, DQL and PG give the same optimal results and are thus shown together.
\label{fig:varyNpulseshape}}
\end{figure}

In Fig.~\ref{fig:varyN} we study a situation where the maximum number of pieces in the control sequence $N$ is given, and the results are shown as the averaged fidelities as functions of $N$. 
Here, the quality of an algorithm is assessed by the averaged fidelity of the state it prepares (as compared to the target state) $\overline{F}$, but not by the computational resources it costs.
For $N\leqslant10$, the Krotov method gives the lowest fidelity, possibly due to the fact that Krotov requires a reasonable level of continuity in the control sequence, and one with a few pieces is unlikely to reach convergence. As $N$ increases, the performance of Krotov is much improved, which has the highest fidelity when $N$ is large ($N\geqslant30$ as seen in the figure). SGD performs better than Krotov for $N\leqslant10$, but worse otherwise, because as $N$ increases, the algorithm has to search over a much larger parameter space. Within the given number of iterations ($N^{\rm iter}=500$ as noted above), it concludes with a lower fidelity. Of course, this result can be improved if more iterations are allowed, and we shall show relevant results in Supplementary Discussion 2. The SGD results at $N=2$ is irregular (thus the cusp at $N=6$), due to the lack of flexibility in the control sequence which makes it difficult to achieve high fidelity with only two steps. 

The fidelity for TQL is higher than SGD and Krotov, but is still lower than that of DQL and PG, indicating the superior ability of deep learning. Nevertheless, we note that the TQL may sometimes fail: it occasionally arrives at a final state which is completely different than the target state. On the other hand, SGD could fail by being trapped at a local minimum, but even in that case it is not drastically different from the optimal solution in terms of the fidelity. This is the reason why the TQL results drop for $N>10$. For larger $N$, the failure rate for TQL is higher (possibly due to the higher dimensionality of the Q-table), and therefore the averaged fidelity is lower. Among all five algorithms, PG is consistently the best. Apart from PG, DQL gives the highest fidelity for $N<30$, but due to its nonzero failure probability, it is outperformed by Krotov for $N>30$. Nevertheless, the effect is moderate and the fidelity is still very close to 1 {($\overline{F}=0.9988$)}.

\begin{figure}
\centering
\includegraphics[width=0.95 \columnwidth]{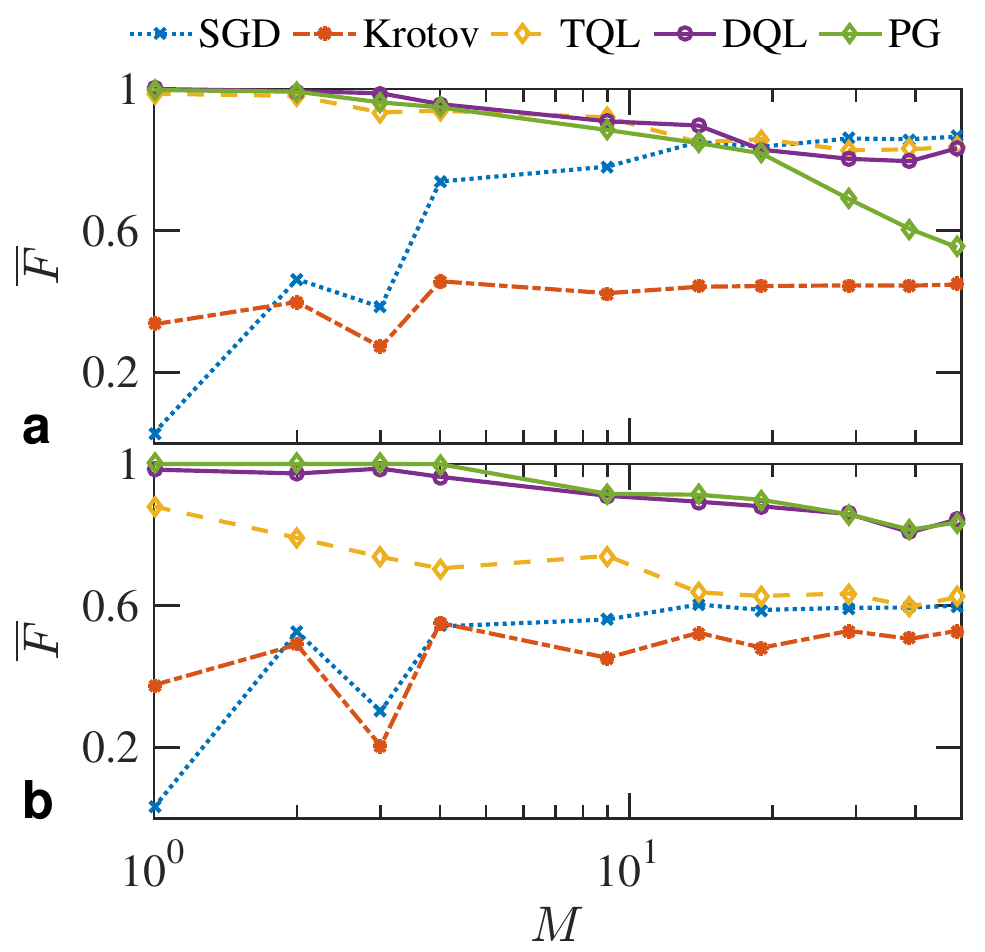}
\caption{ \textbf{Effect of discrete control fields on the averaged fidelity for all five methods considered.} The strength of control field is restricted to $J_i\in[0,1]$, and $M+1$ discrete values (including 0 and 1) are allowed. Panel (a) shows the case of $N=6$ while (b) $N=20$, corresponding to the two vertical dashed lines in Fig.~\ref{fig:varyN}.\label{fig:Jrestrictedgen}}
\end{figure}

To further understand the results shown in Fig.~\ref{fig:varyN}, we take examples from $N=20$ and plot the pulse profiles and the corresponding trajectories on the Bloch sphere in Fig.~{\ref{fig:varyNpulseshape}. We immediately realize that reinforcement learning (TQL, DQL and PG) yield very simple pulse shapes: one only has to keep the control at zero for time $T/2$, and the desired target state ($|1\rangle$) will be achieved. However, to find the result, the algorithm has to somehow realize that one does not have to complete all $N$ pieces, which implies their ability to adaptively generating the control sequence. As can be seen from  Fig.~\ref{fig:varyNpulseshape}a and~\ref{fig:varyNpulseshape}c, SGD and Krotov only search for pulse sequences with exactly $N$ pieces and therefore miss the optimal solution. Their trajectories on the Bloch sphere are much more complex as compared to those of reinforcement learning. In practice, the complex pulse shapes and longer gate times mean that they are difficult to realize in the laboratory, and potentially introduces error to the control procedure (In Supplementary Discussion 4 we provide more details on this issue). From Fig.~\ref{fig:varyNpulseshape} we also notice that reinforcement learning possesses better ability to adaptively sequencing, which is particularly suitable for problems that involve optimization of gate time or speed, such as the quantum speed limit \cite{Zhang.18,Bukov.18}. On the other hand, application of SGD or Krotov to the same problem requires searching over various different $N$ values before an optimal solution can be found, which cost much more resources \cite{Caneva.09,Murphy.10}.

We now study the effect of restrictions on the performances of algorithms. Namely, the control field is bounded between $J_{\min}$ and $J_{\max}$, with $M+1$ allowed values including the bounds. In Fig.~\ref{fig:Jrestrictedgen}, we impose the same restriction $J\in[0,1]$ to all five methods and vary $M$ from $M=1$ to $M=49$. It is interesting to note that the averaged fidelities of three reinforcement learning algorithms decreases with $M$, albeit not considerably. This is because TQL, DQL and PG favor bounded and concrete sets of actions, and more choices will only add burden to the searching process, rendering the algorithms inefficient. Improvements may be made by increasing the number of iterations (cf. Supplementary Discussion 2), and using a larger neural network with stronger representational power.
 For $N=6$ (Fig.~\ref{fig:Jrestrictedgen}a),  TQL and DQL are comparable and have overall the best performance
 except for $M>14$ in which SGD becomes slightly better. On the other hand, $\overline{F}$ for PG drops rapidly for $M\geqslant30$. For $N=20$ (Fig.~\ref{fig:Jrestrictedgen}b), DQL and PG have the best performance, 
 but for large $M$ they are not significantly better than other methods.
More results involving SGD and Krotov are given in Supplementary Discussion 1, from which we conclude that the effect of boundaries in control is much more obvious for Krotov method than SGD, since Krotov performs much larger updates at each iteration. Meanwhile, the effect of discretization (decreasing $M$) are severe for both Krotov and SGD methods, indicating that successful implementations of them depend crucially on the continuity of the problem.

Finally, we note that all results obtained have the target state being $|1\rangle$. Preparing a quantum state other than $|1\rangle$ may have different results, for which an example is presented in Supplementary Discussion 3.  Nevertheless, the overall observation of the pros and cons of the algorithms should remain similar.

\begin{figure}
\centering
\includegraphics[width=0.95 \columnwidth]{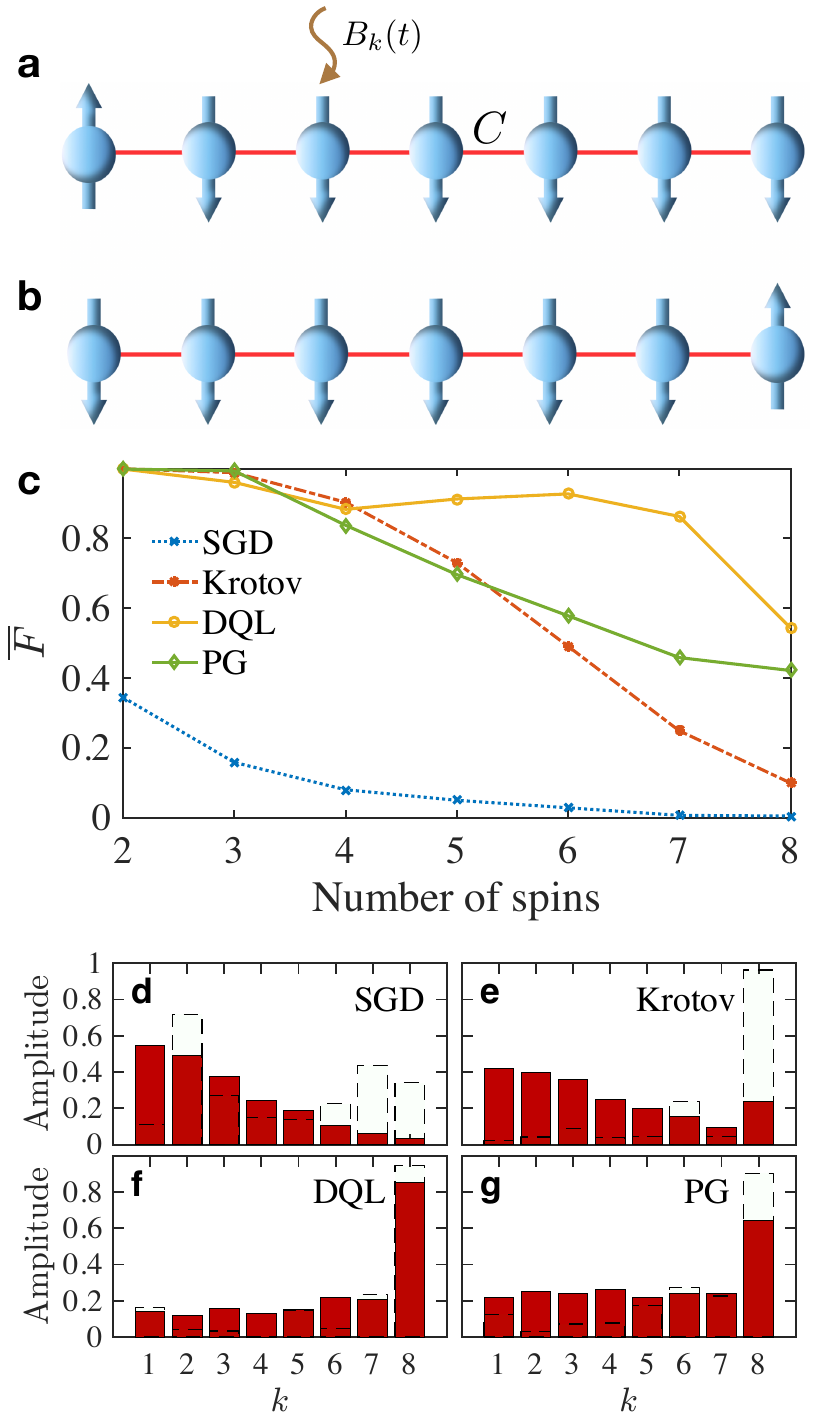}
\caption{ \textbf{Spin transfer as preparation of a multi-qubit state.} (a) The multi-qubit system is initialized with the leftmost spin being up and all others down. (b) The target state has the rightmost spin being up and all others down. (c) The average fidelity versus the number of spins from different algorithms; (d)-(g) The amplitudes (visualization of the final prepared state) at different spins for $K=8$. The red solid bars correspond to results averaged over 100 runs, and the hollow bars enclosed by dashed line shows the results with the highest fidelities.\label{fig:transfer}}
\end{figure}

\begin{table*}
\medskip
\begin{tabular}{cccccc}
\hline
& SGD & Krotov & TQL  & DQL & PG  \\ 
\hline
Performance vs number of time steps $N$ & $\searrow$&$\nearrow$&$\searrow$&$\searrow$&$\star$\\
Ability to adaptively segment&&&$\star$&$\star$&$\star$\\
Discrete operation set  ($M$ small)    & & & &$\star$&$\star$ \\
Continuous operation set ($M$ large) & &$\star$   &&&\\
Scaled-up problems (multi-qubits)&&&&$\star$&\\
\hline
\end{tabular}
\caption{\label{tab:comp} Summary of the performances under different situations. A ``$\star$" indicates that the algorithm performs best, while the arrow ``$\searrow$" (``$\nearrow$") denotes decrease (increase) of the performance versus increase of the variable concerned.
} 
\end{table*}

\subsection{Multi-qubit case}
We now consider a case preparing a multi-qubit state as sketched in Fig.~\ref{fig:transfer}a, b. Our system is described by the following Hamiltonian:
\begin{equation}
H(t)=C\sum^{K-1}_{k=1}\left(S_x^kS_x^{k+1}+S_y^kS_y^{k+1}\right)+\sum_{k=1}^{K}2B_k(t)S^k_z,
\end{equation}
where $K$ is the total number of spins, $S^k_x$, $S^k_y$ and $S^k_z$ are the $k$th spin operator, $C$ describes the constant nearest-neighbor coupling strength (set to be $C=1$), and $B_k(t)$ is the time-dependent local magnetic field applied at the $k$th spin to perform control. This is essentially a task transferring a spin: the system is initialized to a state with the leftmost spin being up and all others down, and the goal is to prepare a state with the rightmost spin being up and all others down. We set the operation time duration to be $T=(K-1)\pi/2$, which is divided to $20$ equal time steps (i.e. $N=20$). The external field is restricted to $B_k(t)/C\in[0,40]$ for SGD and Krotov, and $B_k(t)/C\in\{0,40\}$ for all three reinforcement learning algorithms. Note that TQL fails for $K\ge2$ due to the large size of the Q-table, and is thus excluded in the comparison. 

Fig.~\ref{fig:transfer}c shows the average fidelity versus the number of spins ($K$), after each algorithm is run for 500 iterations. As $K$ increases, the dimensionality of the problem increases and therefore the performances of all algorithms deteriorate. When $K<4$, Krotov, DQL and PG have comparable performances, while SGD has the lowest fidelity. As $K$ increases, $\overline{F}$ for PG and DQL drop much more slowly as compared to Krotov. At $K=8$, we have $\overline{F}=0.0989$ (Krotov), $\overline{F}=0.4214$ (PG), $\overline{F}=0.5433$ (DQL), respectively. Here, we have not assumed a particular form of the control field, so one has to search over a very large space. Specializing the control to certain types would improve performances of the algorithms \cite{Zhang.17}.
 
In order to visualize the final states prepared, we define the amplitude, $A^k$, as the absolute value of the inner product between the final states and the state with the $k$th spin being up while all others being down. A perfect transfer would be that the amplitude is 1 for the rightmost spin and 0 otherwise. Taking $K=8$ as an example, we show how the amplitudes distribute over different spins in Fig.~\ref{fig:transfer}d-g. We compare two different kinds of results: one showing the averaged results over 100 runs (shown as red solid bars), and the other the best result among the 100 runs (hollow bars enclosed by dashed lines). We see that SGD completely fails to prepare the desired state. The best results from Krotov, DQL and PG are comparable, but considering the average over many runs, DQL and PG have better performances. Moreover, the optimal control sequences for different algorithms are provided in Supplementary Table 1-4.

\section{Discussion}

In this paper, we have examined performances of five algorithms: SGD, Krotov, TQL, DQL, and PG, on the problem of quantum state preparation. From the comparison, we can summarize the characteristics of the algorithms under different situations as follows (see also Table.~1).

 Dependence on the maximum number of  pieces in the control sequence, $N$: When all algorithms are executed with the same number of iterations, PG has overall the best performance, but the corresponding fidelity still drops slightly as $N$ increases. In fact, the fidelities from all methods decrease as $N$ increases, except the Krotov method, for which the fidelity increases when $N$ is large. 

 Ability to adaptively segment: During the optimization process, TQL, DQL and PG can adaptively reduce the number of pieces required and can thus find optimal solutions efficiently. SGD and Krotov, on the other hand, always work with a fixed number of $N$ and thus sometimes miss the optimal solution.
 
 Dependence on restricted ranges of the strength of the control field: TQL, DQL and PG naturally work with restricted sets of actions so they perform well when the strength of the control field is restricted. Such restriction reduces the efficiency for both SGD and Krotov method, but the effect is moderate for SGD because its updates on the control field are essentially local. However, the Krotov method makes significant updates during its execution and thus becomes severely compromised when the strength of the control field is restricted. 

 Ability to work with control fields taking $M+1$ discrete values: TQL, DQL and PG again naturally work with discrete values of the control field. In fact, the fidelities from them decrease as the allowed values of the control fields become more continuous ($M$ increases). This problem may be circumvented using more sophisticated algorithms such as  Actor-Critic \cite{Mnih.16,Xu.19}, and the deep deterministic policy gradient method \cite{Lillicrap.15}. SGD is not sensitive to $M$ because it works with a relatively small range of control field and a reasonable discretization is sufficient. The Krotov method, on the other hand, strongly favors continuous problem, i.e.~$M$ being large.

Ability to accommodate scaled-up problems (multiple qubits): Except for TQL, all other algorithms can be straightforwardly generalized to treat quantum control problems with more than one qubit. However, SGD is rather inefficient, and DQL generally outperforms all others for cases considered in this work ($K\le8$).

Moreover, we have found that PG and DQL methods, in general, have the best performances among the five algorithms considered, demonstrating the power of reinforcement learning in conjunction with neural networks in treating complex optimization problems. 

Our direct comparison of different methods may also shed light on how these algorithms can be improved. For example, the Krotov method strongly favors the ``continuous'' problem,   for which TQL, DQL and PG do not perform well. It should be possible that gradients in the Krotov method can be applied in the Q-learning procedures and thereby improves their performances.  We hope that our work has elucidated the effectiveness of reinforcement learning in problems with different types of constraints, and in addition, it may provide hints on how these algorithms can be improved in future studies.

\noindent\textbf{Acknowledgement} This work is supported by the Research Grants Council of the Hong Kong Special Administrative Region, China (Grant Nos.~CityU 21300116, CityU 11303617, CityU 11304018), the National Natural Science Foundation of China (Grant Nos.~11874312, 11604277), the Guangdong Innovative and Entrepreneurial Research Team Program (Grant No.~2016ZT06D348), and Key R\&D Program of Guangdong province (Grant No. 2018B030326001).
 
\noindent\textbf{Code Availability}
The code for all algorithms used in this work is available on GitHub under MIT License (\url{https://github.com/93xiaoming/RL_state_preparation}).

\noindent\textbf{Data Availability}
The data generated during this study are available from the corresponding author upon reasonable request.

\section*{Methods}
In this section, we give a brief description of our implementation of  TQL, DQL and PG in this work. The full algorithms for all methods used in this work are given in Supplementary Method 1.

\subsection{TQL}

For Q-learning, the key ingredients include a set of allowed states $\mathcal{S}$, a set of actions $\mathcal{A}$, and the reward $r$. The state of qubit can be parametrized as
\begin{equation}
    \ket{\psi(\theta,\varphi)} = \pm
    \qty( \cos{\frac{\theta}{2}} \ket{0} + e^{i\varphi} \sin{\frac{\theta}{2}} \ket{1}),
\end{equation}
where $(\theta,\phi)$ corresponds to a point on the Bloch sphere, and a possible global phase of $-1$ has been included. Our set of allowed states is defined as
\begin{equation}
\mathcal{S}\equiv\left\{  \ket{\psi(\theta,\varphi)}|\theta\in s_\theta, \varphi\in s_\varphi \right \},
\end{equation}
where
\begin{equation}
 s_\theta=\left\{\frac{0\pi}{30},\frac{1\pi}{30},\cdots,\frac{29\pi}{30}  \right\} ,\quad s_\varphi= \left\{ \frac{0\pi}{30},\frac{1\pi}{30},\cdots,\frac{59\pi}{30} \right\}.
\end{equation}
We note that this is a discrete set of states, and after each step in the evolution, if the resulting state is not identical to any of the member in the set, it will be assigned as the member that is closest to the state, i.e.~having the maximum fidelity in their overlap.

In the $i$th step of the evolution, the system is at a state $s_i=|\psi_{i}\rangle\in\mathcal{S}$, and the action is given by the evolution operator $a_i=U_i=\exp\{-i H(J_i) dt\}$. All allowed values of the control field $J_i$ therefore form a set of possible actions $\mathcal{A}$. The resulting state $U_i|\psi_{i}\rangle$ after this step is then compared to the target state, and the reward is calculated using the fidelity between the two states as
\begin{eqnarray}
r_i= \left\{
\begin{array}{rcl}
10     &  &  F\in(0.5,0.9],\\    
100   &  &  F\in(0.9,0.999],\\
5000     &  &  F\in(0.999,1],
\end{array} \right. \label{eq:r}
\end{eqnarray}
so that the action that takes the state very close to the target is strongly rewarded. In practice, the agent chooses its action according to the $\epsilon$-greedy algorithm \cite{Sutton}, i.e.~the agent either chooses an action with the largest $Q(s,a)$ with $1-\epsilon$ probability, or with probability $\epsilon$ it randomly chooses an action in the set. The introduction of a nonzero but small $\epsilon$ ensures that the system is not trapped in a poor local minimum. The elements in Q-tables are then updated as:
\begin{equation}
Q(s_{i-1},a_i) \gets Q(s_{i-1},a_i) + \alpha 
    [ r_i + \gamma \mathop{\max}\limits_{a'} Q(s_i,a') - Q(s_{i-1},a_i) ],\label{eq:q1}
\end{equation}
where $a'$ refers to all possible $a_i$ in this step, $\alpha$ is the learning rate, and $\gamma$ is a reward discount to ensure the stability of the algorithm.

\subsection{DQL}

DQL stores the action-value functions with a neural network $\Theta$. We take qubit case as an example. Defining an agent state as 
\begin{equation}
\textbf{s}=\left[\text{Re}\left(\langle0|\psi\rangle\right),\text{Im}\left(\langle0|\psi\rangle\right),\text{Re}\left(\langle1|\psi\rangle\right),\text{Im}\left(\langle1|\psi\rangle\right)\right]^T,
\end{equation}
the network outputs the Q-value for each action $a\in\mathcal{A}$ as $Q(\textbf{s},a;\Theta)$.  We note that in DQL, the discretization of states on the Bloch sphere is no longer necessary and we can deal with states that vary continuously. Otherwise the definitions of the set of actions and reward are the same as those in TQL.

We adopt the double Q-network training approach \cite{Mnih.15}:  two neural networks, the evaluation network $\Theta$ and  the target network $\Theta^-$, are used in training. In the memory we store experiences defined as $e_{i}=(\textbf{s}_{i-1},a_{i},r_{i},\textbf{s}_{i})$. In each training step, an experience is randomly chosen from the memory, and the evaluation network is updated using the outcome derived from the experience.

\subsection{PG} 
Similar to DQL, PG is based on neural networks. With the state $\bm{s}$ as the input vector, the network of PG outputs the probability of choosing each action $\textbf{p}=P(\textbf{s};\Theta)$, where $\textbf{p}=[p_1,p_2,\cdots]^T$. At each time step $t$, the agent chooses its action according to $\textbf{p}$, and stores the total reward it has obtained $v_t=\sum_{i=1}^{t}\gamma^i r_i$. In each iteration, the network is updated in order to increase the total reward. This is done according to the gradient of $\log P(s_t;\Theta) v_t$, the details of which can be found in Supplementary Method 1.

We note that unlike the case for SGD and Krotov, in which the fidelity monotonically increases with more training in most cases, the fidelity output by  TQL, DQL and PG may experience oscillations as the algorithm cannot guarantee optimal solutions in all trials. In this case, one just has to choose outputs which have higher fidelity as the learning outcome.

%

\onecolumngrid
\vspace{1cm}

\begin{center}
{\bf\large Supplementary Materials}
\end{center}
\vspace{0.5cm}

\setcounter{secnumdepth}{3}  
\setcounter{equation}{0}
\setcounter{figure}{0}
\setcounter{table}{0}
\renewcommand{\theequation}{S-\arabic{equation}}
\renewcommand{\thefigure}{S\arabic{figure}}
\renewcommand{\thetable}{S-\Roman{table}}
\renewcommand\figurename{Supplementary Figure}
\renewcommand\tablename{Supplementary Table}

\newcolumntype{M}[1]{>{\centering\arraybackslash}m{#1}}
\newcolumntype{N}{@{}m{0pt}@{}}

\makeatletter \renewcommand\@biblabel[1]{[S#1]} \makeatother


\onecolumngrid

\section*{Supplementary Method 1: Pseudo code for the algorithms}

Here, we provide the pseudo code for all five algorithms used in the main text. 
~\\~\\
\textbf{Algorithm 1: Stochastic Gradient Descendent (SGD).}\\
Set initial guess of $\bm{J}=[J_1,J_2,\cdots,J_N]^T$  randomly\\
\textbf{for}  iteration $=1$, $N^{\rm iter}$\\
 \white{,} \quad Generate a random unit vector $\bm{v}$\\
 \white{,} \quad Set $\bm{J^{+}}=\bm{J}+\alpha \bm{v}, \bm{J^{-}}=\bm{J}-\alpha \bm{v}$\\
 \white{,} \quad Calculate the gradient $\bm{g} = \frac{F(\bm{J^{+}})-F(\bm{J^{-}})}{2 \alpha}$ \\
 \white{,} \quad Update $\bm{J}\leftarrow \bm{J} -\beta \bm{g}$ \\
 \white{,} \quad restrict $\bm{J}$ to the range $[J_{\min},J_{\max}]$ \\
  \textbf{end for} 
~\\~\\
\textbf{Algorithm 2: Krotov algorithm.}\\
Initialize  $\bm{J}$ arbitrarily\\
 Calculate and store $|\psi_i\rangle$ at each step $i$ according to $|\psi_i\rangle=e^{-iH_it}|\psi_{i-1}\rangle$\\
 Set co-state at step $N$ as $\ket{\chi_N} = \ket{\phi}\bra{\phi}\ket{\psi_N}$ \\  
 Calculate and store $|\chi_i\rangle$ at each step $i$ according to $|\chi_{i-1}\rangle=e^{iH_it}|\psi_{i}\rangle$\\  
\textbf{for} iteration $=1$, $N^{\rm iter}$ \\  
\white{,}\quad \textbf{for} $i=1$, $N$ \\
\white{,}\quad\quad Calculate $|\psi_i\rangle$  according to $|\psi_i\rangle=e^{-iH_it}|\psi_{i-1}\rangle$\\
\white{,}\quad\quad Update $ J_i \gets J_i + \Im\langle\chi_i|\partial_{J} H(J_i) |\psi_i\rangle$\\
\white{,}\quad \textbf{end for}\\
\white{,}\quad  Fidelity $F \gets |\ip{\phi}{\psi_N}|^2$\\
\white{,}\quad Set co-state at step $N$ as $\ket{\chi_N} = \ket{\phi}\bra{\phi}\ket{\psi_N}$ \\
\white{,}\quad Calculate and store $|\chi_i\rangle$ at each step $i$ according to $|\chi_{i-1}\rangle=e^{iH_it}|\psi_{i}\rangle$\\
\white{,}\quad Restrict $\bm{J}$ to the range $[J_{\min},J_{\max}]$ \\
\textbf{end for}
~\\~\\
  \textbf{Algorithm 3: Tabular Q-learning (TQL).}\\
 Initialize $Q(s,a)=0$ for all $s \in \mathcal{S}, a \in \mathcal{A}$.\\  
 Initialize the agent state $s_0$ \\    
  \textbf{for}  iteration $=1$, $N^{\rm iter}$\\
 \white{,}\quad \textbf{for} $i=1,N$ \\
 \white{,}\quad\quad With $\epsilon$ probability choose $a_i$ randomly, otherwise  $a_i=\mathop{\arg\max}\limits_{a}Q(s_{i-1},a)$\\
\white{,}\quad\quad Take the action $a_{i}$, evaluate the reward $r_{i}$ and $|\psi_{i}\rangle$\\      
\white{,}\quad\quad Set $s_i$ as the nearest point in $\mathcal{S}$ to $|\psi_{i}\rangle$\\
\white{,}\quad\quad Update $Q(s_{i-1},a_i)$ according to Eq.~(8)\\
\white{,}\quad\quad \textbf{Break} if $1-F<10^{-3}$\\
 \white{,}\quad \textbf{end for}\\
  \textbf{end for}
~\\~\\
  \textbf{Algorithm 4: Deep Q-learning (DQL).}\\
 Initialize memory $R$ as empty\\
Initialize the evaluation network $\Theta$, and target network $\Theta^{-}\leftarrow\Theta$\\
\textbf{for} iteration $=1$, $N^{\rm iter}$\\
\white{,}\quad Initialize $|\psi\rangle=|0\rangle$  and $\bm{s_0}$\\
\white{,}\quad\textbf{for} $i=1$, $N$, \textbf{do}\\
\white{,}\quad\quad With $\epsilon$ probability choose $a_i$ randomly, otherwise  $a_i=\mathop{\arg\max}\limits_{a}Q(\bm{s}_{i-1},a;\Theta)$\\
\white{,}\quad\quad Take the action $a_{i}$, and evaluate the reward $r_{i}$ and state $\bm{s_{i}}$\\
\white{,}\quad\quad Store experience $e_{i}=(\bm{s}_{i-1},a_{i},r_{i},\bm{s}_{i})$ in memory $R$\\
\white{,}\quad\quad \textbf{if} $t$ is divisible by $t_{\rm learn}$\\
\white{,}\quad\quad \quad Sample minibatch of experiences  $e_{k}$\\
\white{,}\quad\quad \quad Set $y_{k}=r_{k}+\gamma \max_{a'}\hat{Q}(\bm{s}_{k},a';\Theta^{-})$\\
\white{,}\quad\quad \quad Update $\Theta$ by minimizing $L=\left[y_{k}-Q(\bm{s}_{k-1},a_{k};\Theta)\right]^{2}$\\
\white{,}\quad\quad \textbf{end if}\\
\white{,}\quad\quad Every $C$ times of learning, set $\Theta^-\leftarrow \Theta$\\
\white{,}\quad\quad \textbf{Break} if $1-F<10^{-3}$\\
\white{,}\quad \textbf{end for} \\
\textbf{end for} 
~\\~\\
  \textbf{Algorithm 5: Policy Gradient (PG).}\\
Initialize the network parameters $\Theta$ arbitrarily\\
\textbf{for} each episode \textbf{do}\\ 
\white{,}\quad \textbf{for}  $t=1$ to  $T-1$ \textbf{do}\\
\white{,}\quad\quad Compute the output of the network $\bm{p}=P(\bm{s};\Theta)$\\
\white{,}\quad\quad Choose action according to the probability distribution $\bm{p}$\\
\white{,}\quad\quad Compute $v_t=\sum_{i=1}^{t}\gamma^{i}r_i$\\
\white{,}\quad\quad Store $s_t, a_t,v_t$\\
\white{,}\quad \textbf{end for}\\
\white{,}\quad Update the network via $\Theta\leftarrow\Theta+\alpha\sum_{t=1}^{T}\nabla_{\Theta}\log P(s_t;\Theta)v_t $\\
 \textbf{end for} 

\vspace{2cm}

\twocolumngrid

\section*{ Supplementary Discussion 1: Effects of bounds and discrete control field for SGD and Krotov methods}

Here, we consider the situation that the control field is bounded for SGD and Krotov, and the results are shown in Fig.~\ref{fig:Jbounded}. Fig.~\ref{fig:Jbounded}a shows the results for the SGD method, with the blue line identical to that in Fig.~2 of the main text (no restriction) and the black one showing results after $J$ is restricted between 0 and 1. We see that imposing a restriction on the available range of the control field does not change the results much, because the search by the SGD algorithm is essentially local: the alteration of $J$ is small in each step and it is unlikely to build up a significant variation of $J$ in the final results. This fact can also be seen from Fig.~3a of the main text: the strength of control field is mostly within the range of $[0,1]$ so that the restriction has minimal effect on the results. 

The situation is different for the Krotov method. As can be seen in Fig.~\ref{fig:Jbounded}b, for $N<30$, the result from the Krotov method with restriction of $J_i\in[0,1]$ (black line) has considerably lower average fidelities than that without restriction (red line, identical to the results shown in Fig.~2 of the main text). This is because the Krotov method makes large updates on the values of the control fields, as can be seen from Fig.~3b of the main text where the magnitude of $J_i$ can be above 20. Restricting the control field to a much narrower range will severely compromise the ability of the algorithm to find solutions with high fidelities. An exception is $N=6$, for which the results with restriction has higher average fidelity.  While the true reason remains unclear, we suspect this is because that the agent happens to have found a relatively good local minimum which outperforms many other cases. We believe that the algorithm succeeds in this particular case but not in general. After all, the averaged fidelity is below 0.6 for both lines, with or without restrictions. For $N>30$, the results without restriction on the range of the control approaches almost one $\left(1-\overline{F}<10^{-7}\right)$, and those with restriction is lower than one but very close (for example, $\overline{F}=0.9822$ for $N=30$). This indicates that having more pieces in the control sequence can greatly help the Krotov algorithm to achieve higher fidelities despite limited strength of control fields.

 \begin{figure}
\centering
\includegraphics[width=0.9 \columnwidth]{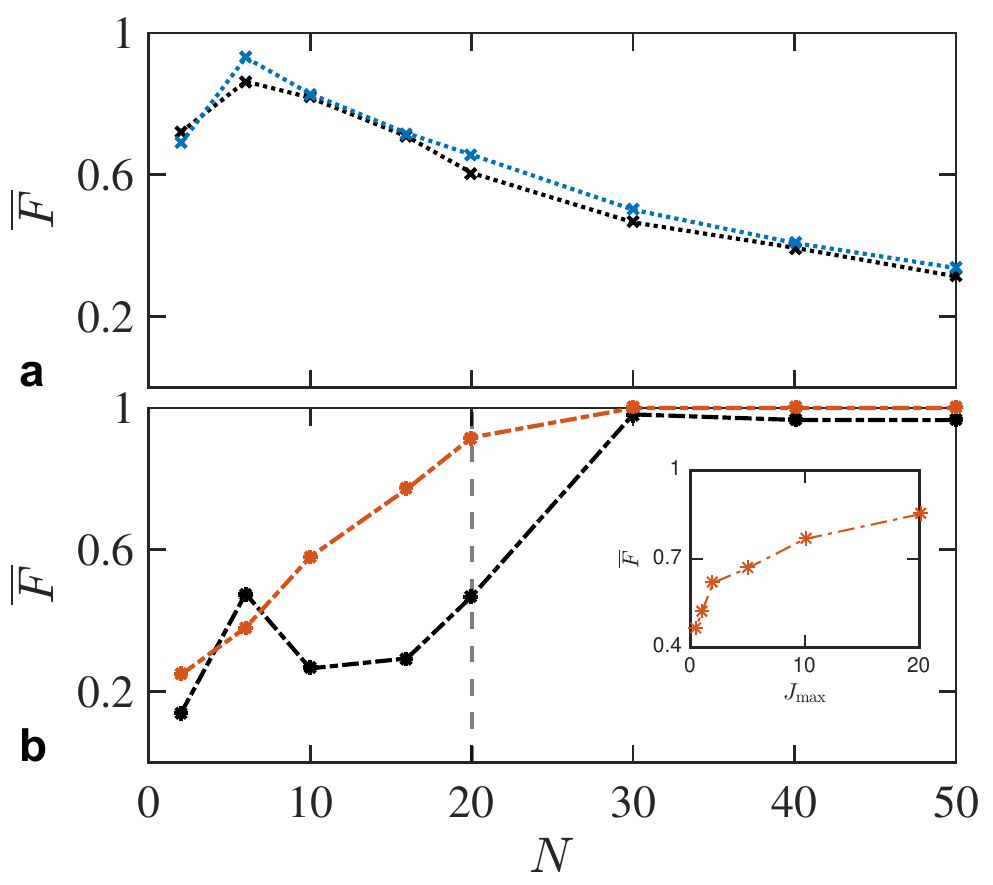}
\caption{ \textbf{Effect of bounds of the control on the average fidelities as functions of $N$ for SGD and Krotov methods.} (a) $\overline{F}$ versus $N$ for the SGD method, without (blue) and with (black) restriction of $J_i\in[0,1]$. (b) Main panel: $\overline{F}$ versus $N$ for the Krotov method, without (red) and with (black) restriction of $J_i\in[0,1]$. Inset: $\overline{F}$ versus $J_{\max}$ for $N=20$, where $J_i$ is restricted to $J_i\in[1-J_{\max},J_{\max}]$.\label{fig:Jbounded}
} 
\end{figure}

\begin{figure}
\centering
\includegraphics[width=0.9 \columnwidth]{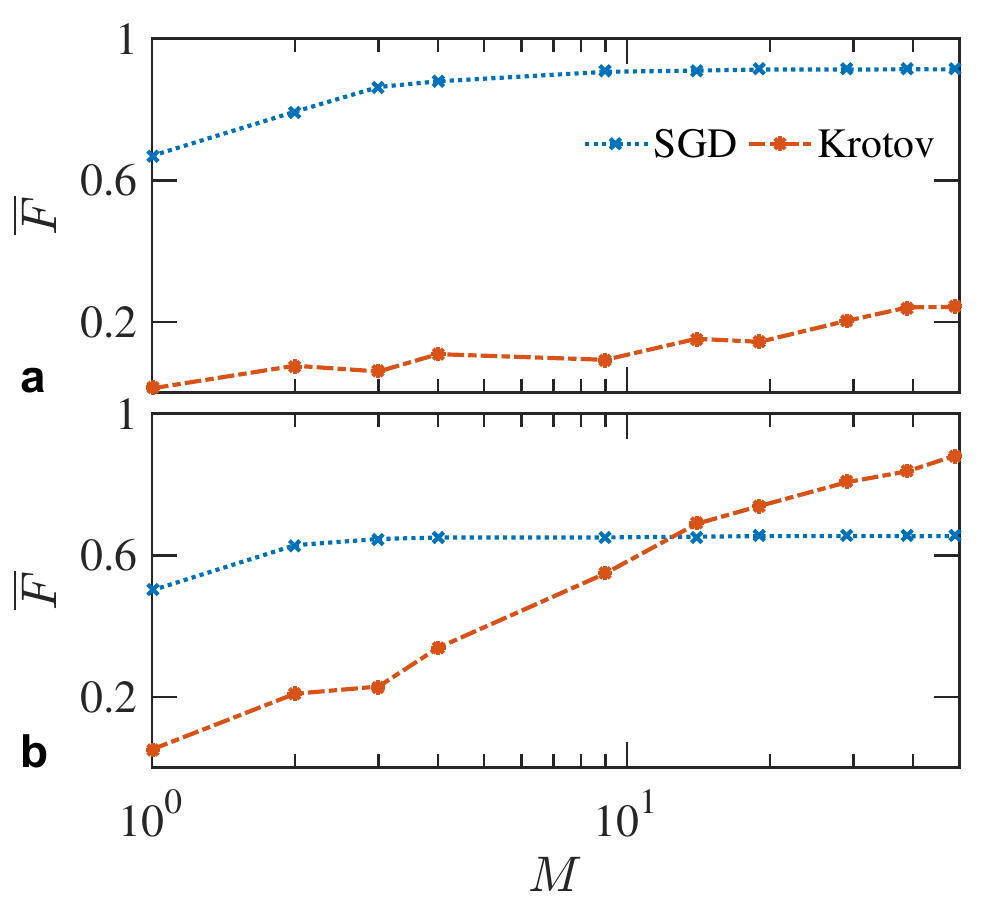}
\caption{\textbf{Effect of discrete control fields on the averaged fidelity for SGD and Krotov methods.} In the calculation, the strength of control field is not specifically restricted, so the $J_{\min}$ and $J_{\max}$  values are determined after the algorithm has run. The values of the control field is then mapped to their respective closest discrete values, with the total number of allowed values including $J_{\min}$ and $J_{\max}$  being $M+1$. Panel (a) shows the case of $N=6$ while (b) $N=20$, corresponding to the two vertical dashed lines in Fig.~2 of the main text.\label{fig:Jdiscretized}
}
\end{figure}
The inset of Fig.~\ref{fig:Jbounded}b gives information on how the two points given by the vertical dashed line at $N=20$ connects when we expand the range of the control field. The bound is given as $1-J_{\max}\le J_{i}\le J_{\max}$. When $J_{\max}$ is increased from 0 to 20, the averaged fidelity from the Krotov method increases from 0.4 to above 0.8. This clearly demonstrates that the range of allowed values of control fields affects the outcome of the Krotov algorithm in a significant way.

We now proceed to consider the effect of discrete control to the averaged fidelities obtained by the algorithms. We start from SGD and Krotov with the range of the control field unrestricted, and the results are shown in Fig.~\ref{fig:Jdiscretized}. In both panels shown, we see that the averaged fidelities from the SGD method first increases for small $M$ but quickly saturate. The insensitivity of the SGD against the discretization of the control field is due to the fact that SGD updates the control field moderately and can find sufficient control field values as desired within a relatively narrow range, even if the values are discretized. This is similar to the reason why the restriction on the range of control field has little effect on the results in Fig.~\ref{fig:Jbounded}a.

On the other hand, the averaged fidelities from the Krotov method increase as functions of $M$, but the increase is much more pronounced for $N=20$ (Fig.~\ref{fig:Jdiscretized}b) than for $N=6$ (Fig.~\ref{fig:Jdiscretized}a). In Fig.~\ref{fig:Jdiscretized}b, the averaged fidelity from Krotov method exceeds that from SGD at around $M+1=15$. The result indicates that successful implementation of the Krotov method depends crucially on the continuity of the problem, in terms of both the number of pieces in the control sequence as well as allowed values of the control field. We also note that at the limit $M\rightarrow\infty$, the extrapolated fidelity values are consistent to results shown in Fig.~2 of the main text, providing a consistency check of our calculations.

\section*{Supplementary Discussion 2: Improving the fidelity with more iterations}

\begin{figure}
\centering
\includegraphics[width=0.9 \columnwidth]{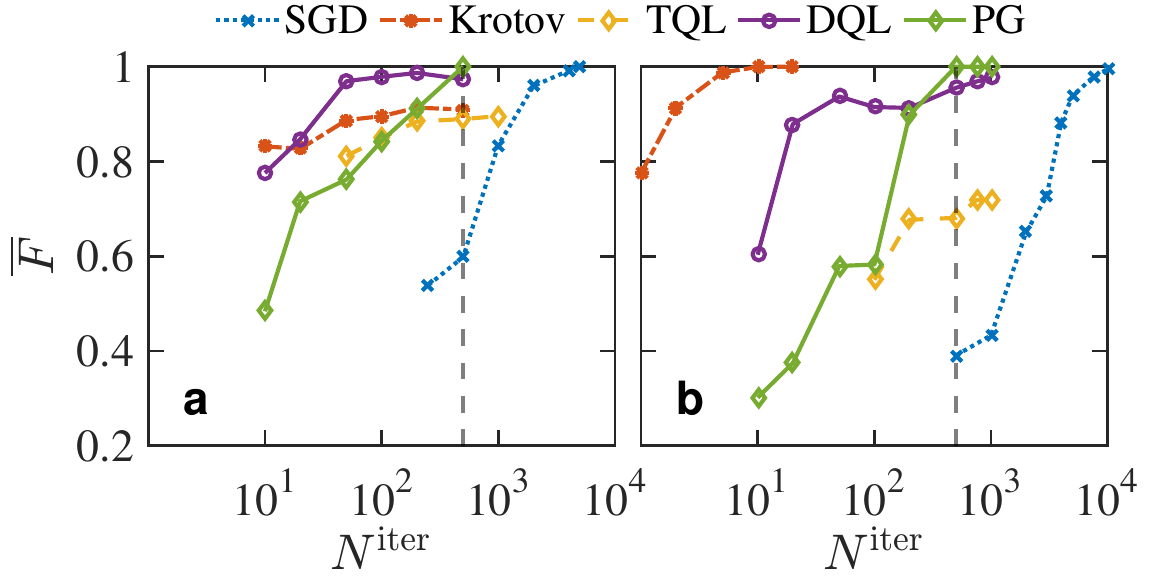}
\caption{ \textbf{Average fidelity versus number of iterations $N^{\rm iter}$}. (a) Maximum number of control pieces $N=20$ and (b) $N=50$. $\overline{F}$ is obtained by averaging over 100 runs.  
The vertical grey dashed lines at $N^{\rm iter}=500$ in both panels correspond to the results shown in Fig.~2 of the main text with the respective $N$ values.}
\label{fig:s1}
\end{figure}

In Fig.~2 of the main text, we have compared five algorithms in terms of the average fidelity versus the maximum number of control pieces. To ensure a fair comparison, all algorithms are requested to stop at $N^{\rm iter}=500$, i.e.~after 500 iterations. Here we show that by allowing more iterations, the fidelity of all algorithms can improve, and the improvement is particularly pronounced for the SGD method. 
Fig.~\ref{fig:s1} shows the average fidelity versus the number of iterations, with all other parameters and constraints the same as those used in Fig.~2 of the main text. Fig.~\ref{fig:s1}a shows the results at $N=20$ for which SGD has a relatively low fidelity (around $0.6$ at $N^{\rm iter}=500$). As the iteration continues, the fidelity of SGD improves substantially, reaching 1 at $N^{\rm iter}\gtrsim500$. On the other hand, the fidelities for other methods do not change much as the number of iteration is increased. Fig.~\ref{fig:s1}b shows the results at $N=50$. We see that results from the Krotov method reaches 1 at $N^{\rm iter}\sim20$, those from DQL improves slightly after  $N^{\rm iter}=50$, but again the increase of fidelity is most pronounced for the SGD method, with the fidelity increasing from 0.4 at $N^{\rm iter}=500$ to 1 at $N^{\rm iter}=10000$. It is also interesting to note that the fidelity output from TQL does not carry a considerable increase, likely because the non-zero failure rate cannot be decreased simply by adding more iterations. We therefore conclude that (1) The result from SGD is most sensitive to the total number of iterations: the fidelity can reach 1 as long as one iterates the algorithm long enough. However, this process could be very resource-consuming compared to other methods that can have high fidelity values for a much smaller number of iterations. (2) The TQL method is most insensitive to more iterations as its intrinsic failure rate cannot be suppressed this way. Adding more iterations will not increase its fidelity output by a notable amount.

\section*{Supplementary Discussion 3: Target state dependence of the learning outcome}

\begin{figure}
\centering
\includegraphics[width=0.9 \columnwidth]{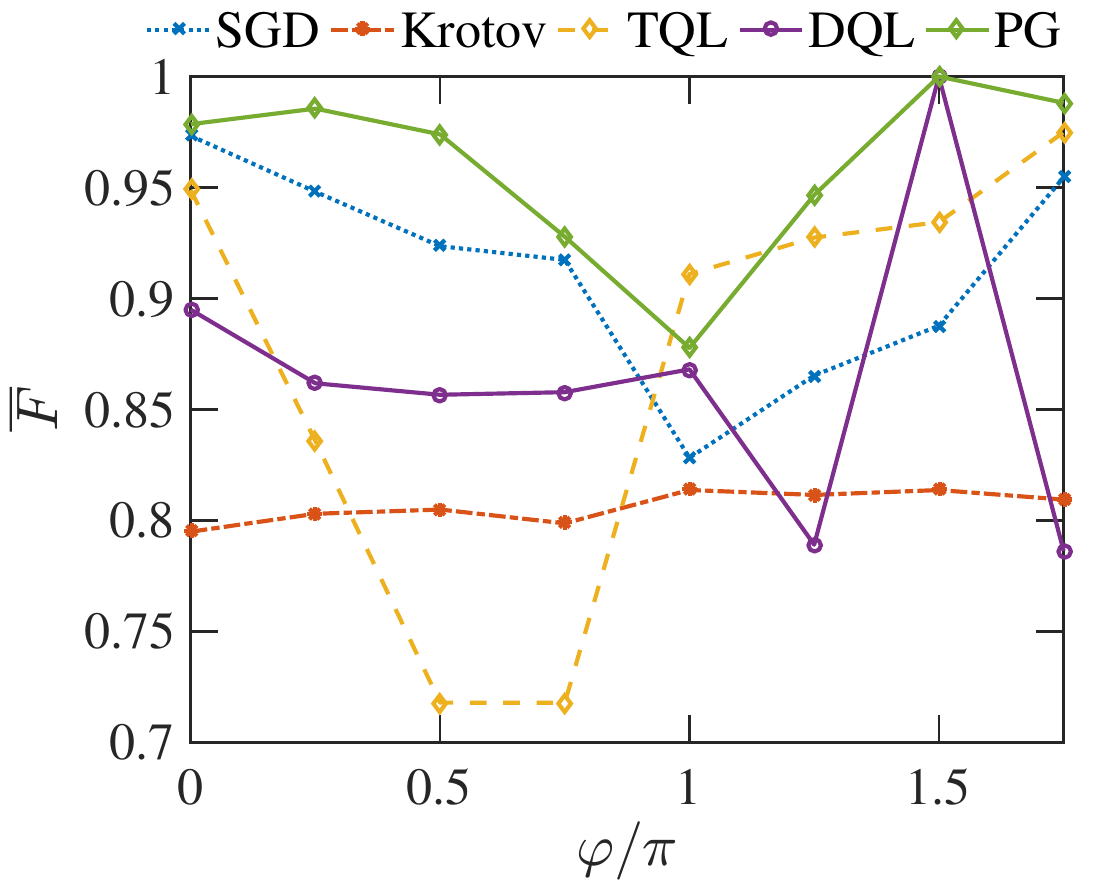}
\caption{\textbf{The dependence of the learning outcome (average fidelity) on the target states.} The target state is defined in Eq.~\eqref{eq:tgs} with $\varphi$ as its sole parameter.} \label{fig:s2}
\end{figure}
In all results shown in the main text, our target state is always $|1\rangle$. In order to provide a more complete picture, we take states on the equator of the Bloch sphere
 \begin{equation}
 |\phi\rangle=\frac{1}{\sqrt{2}}\left(|0\rangle+e^{i\varphi}|1\rangle\right), \tag{S1}
 \label{eq:tgs}
 \end{equation}
as examples of other possible target states. This set of states has one sole parameter $\varphi$ so that one may plot the average fidelity versus $\varphi$ in a figure, which we show in Fig.~\ref{fig:s2} (note that $N^{\rm iter}=500$). First, we see that the Krotov method is not sensitive to the change of states, and the fidelity is maintained at about 0.8 for all target states concerned. All other three methods have outputs that vary a lot with the change of target states. Taking DQL as an example, the average fidelity reaches 1 for $\varphi=1.5\pi$, but is lower than $75\%$ for $\varphi=1.25\pi$. While it may not be meaningful to provide a full explanation, we attribute the variance as the result of a discretized action space, i.e. the allowed actions are limited and cannot cover all points on the Bloch sphere. We note that developments are on-going in order to allow reinforcement  learning to choose action from a continuous set~\cite{Mnih.16,Lillicrap.15}. Their implications on quantum physics warrant further studies.  

\section*{Supplementary Discussion 4: Noise effect} 

\begin{figure}
\centering
\includegraphics[width=0.9 \columnwidth]{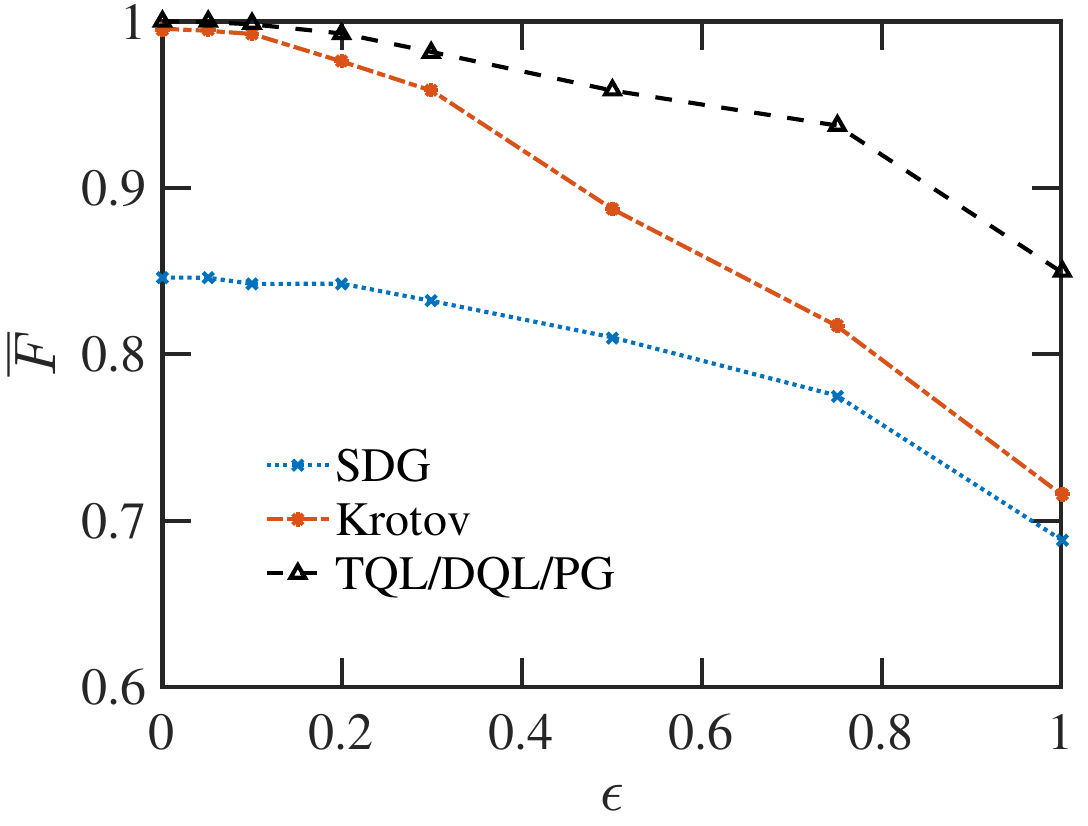}
\caption{ \textbf{Average fidelity versus the noise level in different algorithms.} The target state is $|1\rangle$, $N=20$. Each point of  $\overline{F}$ is an average over 100 runs.}
\label{fig:s3}
\end{figure}

Here, we briefly discuss the robustness of the control sequences found by different algorithms against noises. We first generate the control fields without noises, choosing one for each algorithm that possesses the highest fidelity from 20 runs. Then we feed noise into the evolution via the control field: ${J}_{i}\rightarrow J_{i}+\delta J_i$, where the error term $\delta J_i$ is uniformly drawn from $[-\epsilon,\epsilon]$ with $\epsilon$ being the noise level. Then for each control sequence we produce a set of 100 realizations of noises, and the resulting fidelities are averaged. In Fig.~\ref{fig:s3} we show the results. As is clear from the figure, the average fidelities drop most significantly for Krotov and SGD, but only moderately for TQL and DQL. We believe that it is because the control sequences produced by TQL and DQL are less complex, i.e. having shorter time durations and smaller jumps in the values of the control fields.

\section*{Supplementary Discussion 5: Multiqubit pulse sequence} 
The optimum pulse sequence for the $8$-qubit state transfer problem are shown in Table. S1-S4. The the control sequences here have similar fidelities to the optimum runs shown in Fig. 5(d)-(g) in the main text.
 
\begin{table}[h] 
\begin{center}  
\resizebox{.9\columnwidth}{!}{%
\begin{tabular}{|c|c|c|c|c|c|c|c|c|}  
\hline  
Step &$B_1$&$B_2$ & $B_3$ & $B_4$&$B_5$&$B_6$&$B_7$&$B_8$  \\ \hline  
 $1$&$33.9$&$31.36$&$31.15$&$23.67$&$1.65$&$5.8$&$21.57$&$15.68$\\ \hline 
 $2$&$0.21$&$24.44$&$36.56$&$11.84$&$7.04$&$15.11$&$26.98$&$29.16$\\ \hline 
 $3$&$0.53$&$21.78$&$37.72$&$6.39$&$14.87$&$7.81$&$34.48$&$31.83$\\ \hline 
 $4$&$12.63$&$2.44$&$14.47$&$15.03$&$29.5$&$10.84$&$12.38$&$31.77$\\ \hline 
 $5$&$28.37$&$37.31$&$18.53$&$7.53$&$11.12$&$20.03$&$24.04$&$4.72$\\ \hline 
 $6$&$34.05$&$18.42$&$21.82$&$28.24$&$19.88$&$32.13$&$12.33$&$29.21$\\ \hline 
 $7$&$37.09$&$31.17$&$24.97$&$10.51$&$4.75$&$27.35$&$34.32$&$35.19$\\ \hline 
 $8$&$34.43$&$13.51$&$38.21$&$27.31$&$29.17$&$30.08$&$1.71$&$7.2$\\ \hline 
 $9$&$2.12$&$2.8$&$0$&$6.72$&$12.67$&$20.04$&$18.77$&$35.3$\\ \hline 
 $10$&$29.77$&$36.44$&$37.46$&$21.47$&$7.86$&$0.98$&$21.82$&$1.9$\\ \hline 
 $11$&$25.8$&$8.82$&$1.41$&$0.29$&$7.72$&$10.17$&$18.4$&$28.23$\\ \hline 
 $12$&$20.38$&$37.39$&$20.47$&$9.93$&$26.81$&$19.63$&$35.93$&$14.97$\\ \hline 
 $13$&$4.97$&$22.06$&$38.41$&$40$&$37.41$&$8.08$&$36.03$&$4.24$\\ \hline 
 $14$&$10.71$&$33.98$&$25.2$&$34.33$&$24.52$&$26.29$&$27.03$&$23.94$\\ \hline 
 $15$&$0.3$&$11.93$&$15.24$&$5.52$&$14.06$&$5.94$&$0.54$&$24.58$\\ \hline 
 $16$&$0.84$&$25.63$&$29.83$&$32.55$&$2.03$&$10.75$&$9.13$&$26.51$\\ \hline 
 $17$&$18.12$&$27.29$&$8.65$&$11.68$&$37.27$&$4.66$&$7.72$&$25.27$\\ \hline 
 $18$&$28.43$&$33.35$&$18.71$&$13.84$&$14.58$&$18.06$&$32.33$&$30.47$\\ \hline 
 $19$&$7.41$&$14.36$&$13.81$&$16.79$&$0.88$&$27.72$&$26.64$&$0.11$\\ \hline 
 $20$&$20.23$&$21.58$&$15.56$&$30.26$&$2.91$&$28.78$&$38.92$&$26.01$\\ \hline 
\end{tabular}
}
\caption{Pulse profile for 8-qubit state transfer problem. The result is the one with highest final fidelity among 100 runs of the Krotov algorithm\cite{nt}.}  
\end{center}  
\end{table}

\begin{table}  
\begin{center}  
\resizebox{.9\columnwidth}{!}{%
\begin{tabular}{|c|c|c|c|c|c|c|c|c|}  
\hline  
Step&$B_1$&$B_2$ & $B_3$ & $B_4$&$B_5$&$B_6$&$B_7$&$B_8$  \\ \hline  
 $1$&$21.72$&$10.92$&$18.89$&$29.21$&$12.55$&$6.21$&$5.9$&$10.4$  \\ \hline  
 $2$&$30.38$&$23.33$&$2.54$&$14.49$&$17.5$&$38.92$&$35.58$&$21.58$  \\ \hline  
 $3$&$14.33$&$18.19$&$16.49$&$5.18$&$7.02$&$27.2$&$21.83$&$39.74$  \\ \hline  
 $4$&$29.09$&$16$&$26.28$&$37.61$&$20.51$&$18.65$&$27.32$&$23.52$  \\ \hline  
 $5$&$18.66$&$8.07$&$36.99$&$15.61$&$3.8$&$20.36$&$5.54$&$6.89$  \\ \hline  
 $6$&$33.59$&$15.99$&$3.08$&$8.13$&$20.27$&$4.34$&$22.21$&$37.9$  \\ \hline  
 $7$&$2.71$&$3.72$&$2.86$&$37.55$&$1.54$&$31.14$&$8.68$&$19.23$  \\ \hline  
 $8$&$33.23$&$28.24$&$29.05$&$1.45$&$33.59$&$17.74$&$33.82$&$35.73$  \\ \hline  
 $9$&$15.95$&$32.9$&$9.55$&$12.14$&$4.82$&$0.39$&$18.7$&$37.08$  \\ \hline  
 $10$&$38.96$&$39.31$&$15.11$&$15.45$&$21.56$&$6.29$&$4.26$&$4.75$  \\ \hline  
 $11$&$15.54$&$38.26$&$23.76$&$14.82$&$0.81$&$37.93$&$15.23$&$12.11$  \\ \hline  
 $12$&$3.18$&$22.49$&$39.74$&$30.15$&$27.02$&$1.57$&$27.61$&$0.58$  \\ \hline  
 $13$&$24.28$&$33.64$&$29.41$&$26.82$&$29.1$&$11.96$&$14.37$&$19.22$  \\ \hline  
 $14$&$20.57$&$2.68$&$24.79$&$39.48$&$8.48$&$8.06$&$23.21$&$39.6$  \\ \hline  
 $15$&$29.03$&$29.28$&$37.98$&$31.63$&$12.09$&$24.96$&$31.67$&$39.74$  \\ \hline  
 $16$&$37.28$&$17.21$&$5.65$&$4.65$&$12.69$&$28.18$&$29.02$&$15.14$  \\ \hline  
 $17$&$4.48$&$2.19$&$22.02$&$19.95$&$24.55$&$24.57$&$17.49$&$15.59$  \\ \hline  
 $18$&$19.8$&$4.43$&$12$&$36.86$&$36.75$&$3.14$&$20.75$&$12.66$  \\ \hline  
 $19$&$9.3$&$19.9$&$30.61$&$24.88$&$29.13$&$24.43$&$36.7$&$12.03$  \\ \hline  
 $20$&$11.02$&$37.63$&$38.67$&$16.97$&$14.76$&$1.47$&$23.02$&$25.28$  \\ \hline  
\end{tabular}  
}
\caption{Pulse profile for 8-qubit state transfer problem. The result is the one with highest final fidelity among 100 runs of the SGD algorithm\cite{nt}.}  
\end{center}  
\end{table}  

\begin{table}
\resizebox{.75\columnwidth}{!}{%
\begin{tabular}{|c|c|c|c|c|c|c|c|c|}  
\hline  
Step&$B_1$&$B_2$ & $B_3$ & $B_4$&$B_5$&$B_6$&$B_7$&$B_8$  \\ \hline  
 $1$&$0$&$0$&$0$&$0$&$0$&$0$&$40$&$0$\\ \hline  
 $2$&$0$&$0$&$0$&$0$&$0$&$0$&$40$&$0$\\ \hline  
 $3$&$0$&$40$&$40$&$40$&$40$&$40$&$40$&$0$\\ \hline  
 $4$&$40$&$40$&$0$&$40$&$40$&$40$&$40$&$40$\\ \hline  
 $5$&$40$&$40$&$0$&$40$&$40$&$40$&$40$&$40$\\ \hline  
 $6$&$0$&$40$&$0$&$0$&$40$&$0$&$0$&$40$\\ \hline  
 $7$&$0$&$40$&$0$&$0$&$40$&$0$&$0$&$40$\\ \hline  
 $8$&$0$&$40$&$0$&$0$&$40$&$0$&$0$&$40$\\ \hline  
 $9$&$0$&$40$&$0$&$0$&$40$&$0$&$0$&$40$\\ \hline  
 $10$&$0$&$40$&$0$&$0$&$40$&$0$&$0$&$40$\\ \hline  
 $11$&$0$&$40$&$0$&$0$&$40$&$0$&$0$&$40$\\ \hline  
 $12$&$40$&$0$&$40$&$40$&$0$&$0$&$40$&$0$\\ \hline  
 $13$&$0$&$40$&$0$&$0$&$40$&$0$&$0$&$40$\\ \hline  
 $14$&$0$&$40$&$0$&$0$&$40$&$0$&$0$&$40$\\ \hline  
 $15$&$0$&$40$&$0$&$0$&$40$&$0$&$0$&$40$\\ \hline  
 $16$&$0$&$40$&$0$&$0$&$40$&$0$&$0$&$40$\\ \hline  
 $17$&$0$&$40$&$0$&$0$&$40$&$0$&$0$&$40$\\ \hline  
 $18$&$0$&$40$&$0$&$0$&$40$&$0$&$0$&$40$\\ \hline  
 $19$&$0$&$40$&$0$&$0$&$40$&$0$&$0$&$40$\\ \hline  
 $20$&$0$&$40$&$0$&$0$&$40$&$0$&$0$&$40$\\ \hline  
\end{tabular}  
}
\caption{Pulse profile for 8-qubit state transfer problem. The result is the one with highest final fidelity among 100 runs of the DQL algorithm\cite{nt}.}  
\end{table} 
$\\\\\\\\\\\\\\\\\\\\$
\begin{table}
\resizebox{0.75 \columnwidth}{!}{%
\begin{tabular}{|c|c|c|c|c|c|c|c|c|}  
\hline  
Step&$B_1$&$B_2$ & $B_3$ & $B_4$&$B_5$&$B_6$&$B_7$&$B_8$  \\ \hline  
 $1$&$0$&$0$&$40$&$40$&$0$&$40$&$40$&$40$\\ \hline
 $2$&$0$&$40$&$40$&$0$&$40$&$40$&$40$&$40$\\ \hline
 $3$&$0$&$0$&$40$&$40$&$0$&$40$&$40$&$40$\\ \hline
 $4$&$0$&$40$&$40$&$40$&$0$&$0$&$0$&$0$\\ \hline
 $5$&$40$&$0$&$40$&$0$&$0$&$0$&$40$&$0$\\ \hline
 $6$&$0$&$0$&$0$&$40$&$0$&$40$&$40$&$0$\\ \hline
 $7$&$0$&$0$&$40$&$0$&$0$&$0$&$0$&$40$\\ \hline
 $8$&$0$&$0$&$40$&$0$&$40$&$0$&$0$&$0$\\ \hline
 $9$&$0$&$0$&$40$&$0$&$40$&$0$&$0$&$0$\\ \hline
 $10$&$0$&$0$&$40$&$40$&$0$&$40$&$40$&$40$\\ \hline
 $11$&$40$&$0$&$0$&$0$&$0$&$0$&$40$&$0$\\ \hline
 $12$&$0$&$40$&$40$&$40$&$0$&$0$&$0$&$0$\\ \hline
 $13$&$40$&$0$&$40$&$40$&$0$&$0$&$40$&$40$\\ \hline
 $14$&$0$&$40$&$0$&$40$&$0$&$40$&$0$&$0$\\ \hline
 $15$&$40$&$0$&$40$&$40$&$40$&$0$&$0$&$40$\\ \hline
 $16$&$0$&$40$&$0$&$0$&$0$&$0$&$40$&$40$\\ \hline
 $17$&$0$&$40$&$40$&$0$&$40$&$40$&$40$&$40$\\ \hline
 $18$&$40$&$0$&$40$&$40$&$40$&$40$&$0$&$40$\\ \hline
 $19$&$40$&$0$&$40$&$40$&$40$&$0$&$0$&$40$\\ \hline
 $20$&$0$&$0$&$0$&$0$&$0$&$40$&$40$&$0$\\ \hline
\end{tabular} 
} 
\caption{Pulse profile for 8-qubit state transfer problem. The result is the one with highest final fidelity among 100 runs of the PG algorithm\cite{nt}.}  
\end{table}  

%

\begin{thebibliography}{51}
\bibitem [{nt()}]{nt}%
  \BibitemOpen
  \href@noop {} {}\bibinfo {note} {Sequences generated here have similar
  fidelities to those in Fig. 5 in the main text.}\BibitemShut {Stop}%
\end{thebibliography}

\end{document}